\documentclass[preprint,12pt]{elsarticle}

\usepackage{amssymb}
\usepackage{amsmath}
\usepackage{gensymb}
\usepackage{algpseudocode}
\usepackage[linesnumbered,ruled,vlined]{algorithm2e} 
\usepackage{graphicx}
\usepackage{textcomp}
\usepackage[table]{xcolor} 
\usepackage{booktabs} 
\usepackage{tabularx} 
\usepackage{xcolor}
\usepackage{array}
\usepackage{rotating}

\journal{Physical Communication}

\begin{document}
\definecolor{customgreen1}{RGB}{178, 223, 219} 
\definecolor{customgreen2}{RGB}{163, 217, 166} 
\definecolor{customgreen3}{RGB}{218, 235, 212} 
\definecolor{customgreen4}{RGB}{150, 200, 180}

\begin{frontmatter}



\title{Proactive Blockage Prediction for UAV-assisted Handover in Future Wireless Network}


\author[I. Ahmad]{Iftikhar Ahmad}
\author[A. Khan]{Ahsan Raza Khan}
\author[A. Jabbar]{Abdul Jabbar}
\author[M. Alquraan]{Muhammad Alquraan}
\author[L. Mohjazi]{Lina Mohjazi}
\author[M. Ur Rehman]{Masood Ur Rehman}
\author[M. A. Imran]{Muhammad Ali Imran}
\author[A. Zoha]{Ahmed Zoha}
\author[S. Hussain]{Sajjad Hussain\corref{cor1}}
\cortext[cor1]{Corresponding author: Sajjad Hussain (sajjad.hussain@glasgow.ac.uk)}

\affiliation{organization={James Watt School of Engineering, University of Glasgow}, 
            city={Glasgow},
            postcode={G128QQ},
            country={United Kingdom}}

\begin{abstract}
The future wireless communication applications demand seamless connectivity, higher throughput, and low latency, for which the millimeter-wave (mmWave) band is considered a potential technology. Nevertheless, line-of-sight (LoS) is often mandatory for mmWave band communication, and it renders these waves sensitive to sudden changes in the environment. Therefore, it is necessary to maintain the LoS link for a reliable connection. One such technique to maintain LoS is using proactive handover (HO). However, proactive HO is challenging, requiring continuous information about the surrounding wireless network to anticipate potential blockage. This paper presents a proactive blockage prediction mechanism where an unmanned aerial vehicle (UAV) is used as the base station for HO. The proposed scheme uses computer vision (CV) to obtain potential blocking objects, user speed, and location. To assess the effectiveness of the proposed scheme, the system is evaluated using a publicly available dataset for blockage prediction. The study integrates scenarios from Vision-based Wireless (ViWi) and UAV channel modeling, generating wireless data samples relevant to UAVs. The antenna modeling on the UAV end incorporates a polarization-matched scenario to optimize signal reception. The results demonstrate that UAV-assisted Handover not only ensures seamless connectivity but also enhances overall network performance by $20 \%$. This research contributes to the advancement of proactive blockage mitigation strategies in wireless networks, showcasing the potential of UAVs as dynamic and adaptable base stations.
\end{abstract}




\end{frontmatter}



\section{Introduction}
Millimetre-wave (mmWave) and terahertz (THz) communication technologies are envisioned as the potential candidates to cater for the ever-increasing data rates demands \cite{j1}. These technologies offer massive connectivity, ultra-reliable low latency communication (URLLC), and higher bandwidth for sophisticated applications like smart healthcare, industry 4.0, holographic telepresence, virtual and augmented reality (VR/AR) and self-driving cars \cite{j2, jabbar2022millimeter}. Furthermore, the shift to higher frequency bands alters the paradigm of forthcoming wireless networks by emphasizing small coverage cells, giving rise to the concept of ultra-dense networks (UDNs) \cite{kamel2016ultra}. In future networks, mmWave and THz multiarray antennas are exploited, providing beamforming capabilities that concentrate the radio signal power onto the receiving device through line-of-sight (LoS) communication \cite{9926150}.
Despite various merits, the mmWaves and THz communication are prone to higher penetration losses, difficulty supporting mobility, and are very sensitive to blockages. As an illustration, a link budget experiences a power loss of 20 dB or greater when the connection is obstructed by obstacles like human bodies or vehicles \cite{yamamoto2008path}. Therefore, these technologies rely heavily on a line-of-sight (LoS) communication link between the base station (BS) and the intended user \cite{j1, j3}. 

The challenge of link blockage can be overcome by developing a sense of wireless network surroundings to anticipate the potential blockage. The traditional approach to deal with this challenge is the combination of machine learning and wireless sensor data (e.g., channel, received signal strength). Recent studies provided both reactive and proactive blockage prediction using wireless sensor data \cite{j4}. The reactive blockage prediction does not satisfy the low-latency requirement, whereas the proactive blockage prediction is still emerging which requires thorough investigation. The future blockage prediction enables the wireless system to make informed decisions, such as proactive handover (HO), to maintain seamless connectivity. 

Unmanned aerial vehicles (UAVs) are considered valuable service enablers for smart city applications, the healthcare domain, real-time surveillance and monitoring, disaster management, and wireless communication \cite{charan2022towards}. Due to their airborne positioning and capability to deploy on demand in specified areas, UAVs may be seen as flying BSs that can be utilised for conducting massive MIMO, 3D networked MIMO, and mmWave communications \cite{huang2021massive, zhang2019research, garcia2019essential}. As a result, Unmanned Aerial Vehicles (UAVs) are commonly employed as aerial base stations or relays to enhance network capacity and offer more adaptable coverage \cite{ahmad2022uav}. Inspired by the promising prospects of UAV-assisted proactive communication, this paper presents a proactive blockage prediction for HO using UAV as BS, leveraging vision and wireless data. In the HO mechanism, the users experiencing performance degradation are shifted to other BS with higher signal strength. However, a successful HO needs surrounding information and proactive blockage prediction. Furthermore, false or frequent HO leads to more delays, throughput loss, and low quality of service (QoS). Therefore, blockage prediction is an active area of research to find novel and scalable solutions, ensuring the reliability and performance of the wireless network.

\begin{figure*}[!t]
	\centerline{\includegraphics[width=\linewidth]{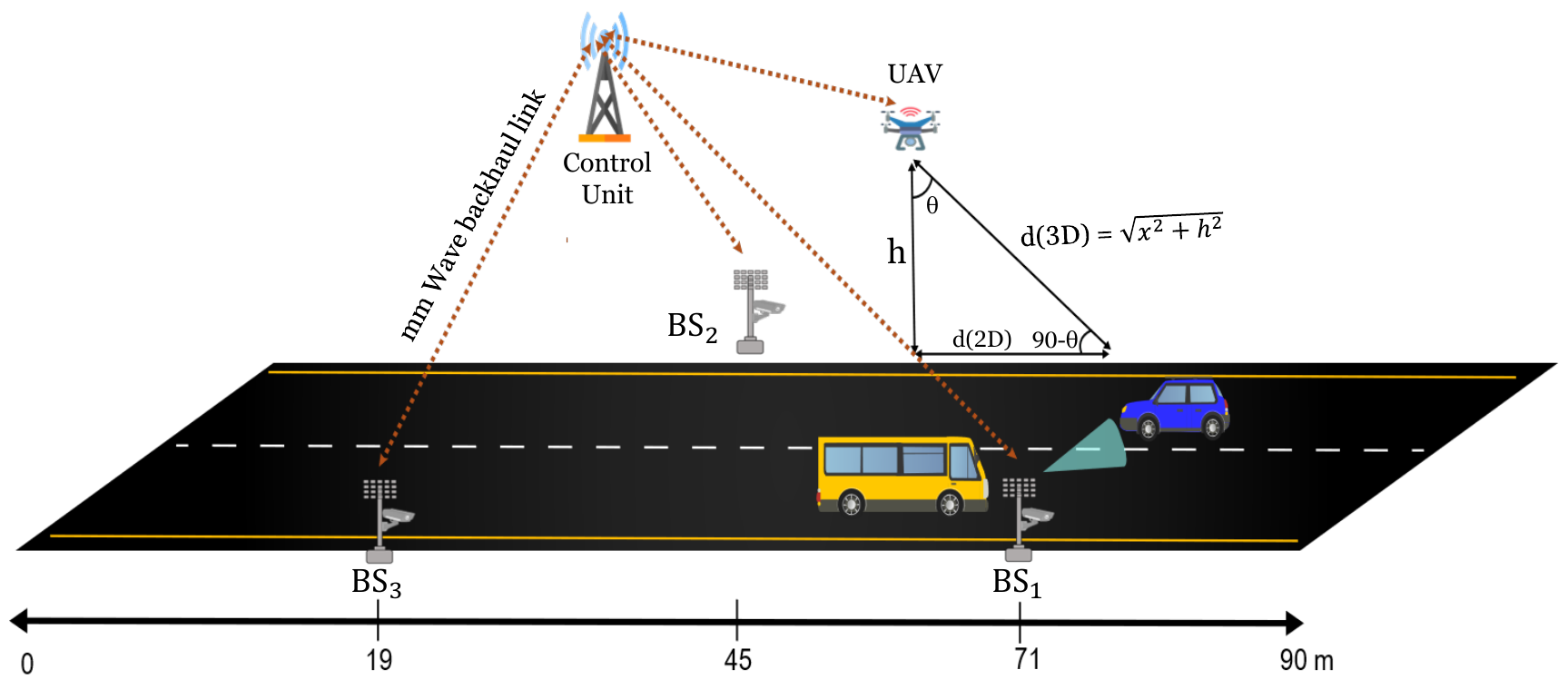}}
	\caption{UAV-based vision-aided wireless communication system having central control, three BS equipped with RGB camera, and a UAV base station providing coverage to blocked area.}
	\label{sm}
\end{figure*}
\subsection{Related Works} 

Despite numerous advantages of mmWaves and THz communication, using these high-frequency bands introduces many challenges, including sensitivity to LoS blockage and higher training overhead \cite{U.8, rappaport2019wireless}. As a result, a significant drop in QoS occurs due to blockages. One of the possible solutions to address this challenge is multi-connectivity, where users connect with multiple BS simultaneously \cite{j5}. In this approach, each BS measures the link quality of the connected user and feeds its measurement report to central control, which then provides the optimal scheduling mechanism to maintain the quality of the communication link. However, the multi-connectivity algorithms are usually reactive and triggered in response to blockages. Furthermore, they also increase the scheduling overhead and introduce undesirable latency \cite{j1}. 

The other possible solution to maintain seamless connectivity in high-frequency communication is proactive HO. However, the HO mechanism requires prior knowledge of surroundings and information about real-timelink blockages. Therefore, various studies have been conducted to solve the problem of link blockage prediction. Prediction approaches, particularly those centered on beam prediction tasks, have seen a growing interest in leveraging machine learning (ML) in recent years. These solutions primarily concentrate on utilizing additional information to improve understanding of the wireless environment \cite{charan2022towards}. For instance, the studies proposed machine learning (ML) techniques to predict the link blockages using wireless sensor data (RSS, and channel characteristics) \cite{j5, j7}. However, these studies perform reactive actions that impact the performance of the communication system. Inspired by the proficiency of machine learning models the authors in \cite{j8}, proposed a proactive blockage prediction algorithm for mmWave communication. The proposed techniques used the sequence of beamforming to train the Gated Recurrent Unit (GRU) that can predict the link blockages. Although, this technique is very simple and effective but very sensitive to sudden changes in the channel due to a single data modality.  Authors in \cite{alrabeiah2020deep}, recommended using sub-6GHz channels to predict future link blockages to foresee future dynamic or mobile blockages. Additionally, in the field of mmWave and THz networks, predicting signal blockages remains a significant challenge due to the dynamic nature of environmental changes. A recent study \cite{khan2024semantic} proposes a Semantic-Aware Federated Blockage Prediction (SFBP) framework to address this issue by integrating computer vision and distributed on-device learning. This framework utilizes semantic information extracted from images to enhance blockage prediction accuracy.

The use of multi-modal data and fusion of deep learning (DL) and computer vision (CV) is an emerging trend to solve the challenge of link blockage in high-frequency communication. The fusion of multi-modal data allows the wireless system to have a sense of the surrounding environment that is envisioned to play a vital role in future wireless communication, especially in blockage prediction, HO, and network resource allocation. Some of the preliminary studies leveraging the multi-modal CV and DL are presented in the literature. For instance,   a vision-aided proactive handover algorithm is proposed in \cite{U.6} using depth images and wireless data. A DL model is trained using multi-modal data to learn the relationship between the depth images and measured throughput to estimate future link quality and provide timely information for optimal HO. Similarly, the study in \cite{U.8}, utilized the RGB images and beamforming vector for training a DL for proactive blockage prediction. The development of machine learning-based solutions is on the rise, driven by the limitations of traditional systems in effectively addressing multi-user settings with high mobility. These systems are engineered to leverage prior research findings and diverse sensing information, such as user position \cite{morais2023position}, camera/visual images \cite{charan2022vision}, LiDAR data \cite{jiang2022lidar}, and radar data \cite{demirhan2022radar}. The authors in \cite{ahmad2023vision} introduced an innovative approach that combines computer vision (CV) and ensemble learning, specifically through stacking. This approach integrates multi-modal vision sensing and positional data to achieve precise estimations of UAV positions and orientations. The results demonstrate that the proposed method accurately predicts K-beams and significantly improves the overall performance of mmWave drone communication networks.
According to the research studies available in the literature, it is evident that utilizing multi-modal data will significantly improve the performance of the wireless communication system. However, most techniques using multi-modal data predict the potential blockage but lack the action needed to maintain seamless connectivity.

\subsection{Motivation and Contributions} 

As discussed earlier, using mmWaves introduces various challenges. Nevertheless, the link blockage is one of the key bottlenecks for high-frequency communication. The future wireless communication requires higher QoS with seamless connectivity to serve real-time applications. Therefore, proactive HO is one of the promising solutions to maintain connectivity by shifting the user to another LoS link. However, proactive HO requires prior knowledge of link blockage to perform timely action. Multi-modal technique along with CV and DL is envisaged to assist the smooth operation of wireless systems in this regard. 

Our previous study proposed a novel CV-assisted HO mechanism using multi-modal vision (RGB images) and wireless data (RSSI) \cite{9926150}. The proposed scheme introduced a new HO event termed a blockage event (BLK), which indicates the presence and location of potential blockage when the user is in the field-of-view (FoV) of the vision sensor. The combination of CV and multi-variate regression, a metric for HO i.e., time to block $(T_{blk})$, is obtained, and actions are taken to maintain the seamless connectivity. However, shifting the user from one BS to another degrades the signal strength caused by the path loss.

In this paper, we address this challenge by proposing a solution for the UAV-assisted HO process using a combination of multi-modal vision (RGB images) and wireless data represented by the received signal strength indicator (RSSI). The system employs Object Detection and Localization (ODL) to identify the user's presence, potential blockage, user location, and distance from blocking objects. Additionally, a neural network predicts the time required for handover, denoted as $T_{blk}$, and appropriate corrective measures are implemented if sufficient time for handover is available.

To extend coverage to the blocked area, a UAV positioned at a specific height acts as a base station. When a blockage is detected ($BLK$ triggered), the proposed algorithm initiates a handover request, seamlessly transitioning the user to the UAV. This ensures continuous connectivity and enhances signal strength. The main contributions of this paper are highlighted as:
\begin{itemize}
    \item The paper proposes UAV-assisted HO utilizing CV and ML to address the issue of link blockages for high-frequency communication. Multi-modal data (vision and wireless) is employed for proactive blockage prediction to facilitate successful HO with minimal performance degradation. The integration of CV with multi-modal data enhances the network's awareness of its surroundings, leading to improved blockage prediction.
    \item We developed the channel modeling for UAV-based BS and provided an analytical model for dipole antennas for UAV-to-ground communication. Furthermore, a detailed analysis is performed to study the impact of path loss on RSSI by placing the UAV at different heights.
    \item Finally, a comparative analysis of UAV-assisted and non-UAV HO validates the effectiveness of the proposed scheme. The results demonstrate a 20\% improvement in RSSI using UAV-assisted HO.
\end{itemize}
\subsection{Paper Organisation}
The rest of the paper is organized as follows: Section II explains the system model and UAV channel model adopted in this work. Section III describes the UAV-assisted HO mechanism. Section IV describes the simulation setup and discussion of the results. Section V concludes the paper with a brief outlook of future research direction. 

\section{System Model}
This work aims to use vision and wireless data with DL to predict the potential blockage and perform a proactive handover. The idea is to use different technologies like CV, DL, and UAV-assisted communication to aid wireless communication in a higher-frequency band. The following subsections provide a detailed description of UAV-based vision-aided wireless communication. 

\subsection{Scenario Description} 
A high-frequency wireless communication system is considered to cover an urban street of $90$m x $15$m in length and width, respectively, as illustrated in Fig. \ref{sm}. There are three small base stations (SBS), central control, and a UAV placed at a particular height ``h", covering the blockage area. The SBSs are equipped with a uniform linear array (ULA) antenna with $M$ elements, using beamforming techniques to create the line-of-sight (LOS) link that can achieve high signal strength. The communication system uses orthogonal frequency division multiplexing (OFDM) operating at the unlicensed 60 GHz frequency band. A codebook-based beamforming at 60 GHz multi-antenna OFDM systems is reported in \cite{U1}. Furthermore, each SBS is also equipped with standard RGB cameras to monitor the environment and capture visual information to predict potential blockages. For efficient handover (HO), the UAV is an alternative to SBS and provides coverage to blocked areas. For the simplicity of the scenarios, we consider a single moving user (car), a stationary blockage object (bus), three SBSs covering the entire street, and a static UAV placed above the blockage area as depicted in Fig. \ref{sm}. The sensors at the SBSs capture the vision and wireless information of the surroundings and share it with the control unit (CU) via a 10 Gbps point-to-point mmWave backhaul link \cite{U3}. The CU is the brain of this system, which collects and processes the related information to train the ML model for proactive blockage prediction. Furthermore, once the ML model is trained, the CU ensures a smooth HO using the real-time data.

\subsection{UAV Channel Modelling} \label{uav_channel}
This work considers the LOS communication to investigate the properties of air-to-ground channels for UAVs. Once the moving car enters the scenario, the vision sensors track the user and look for potential blockage. The UAV is placed at the optimal position if there is a blocking object, as shown in Fig. \ref{sm}. In the case of UAVs, an omnidirectional antenna is preferred for air-to-ground communication over directional antennas to avoid any possible transmitter/receiver (Tx/Rx) alignment issues due to the high-speed movements of the cars (user). Directional antennas, however, can be employed if both Tx and Rx are static or show small displacements with real-time reconfigurability in antenna patterns.

For an unobstructed UAV scenario, the antenna gain for the LOS component in the elevation plane significantly influences the co-polarised antenna's received power. We present an analytical path loss model based on the antenna gain in the elevation plane to address this challenge. Different UAV altitudes, antenna orientations, and the impact of elevation angle on the received power are studied.

\begin{figure}[!t]
 \centering
     \includegraphics[width=7cm, keepaspectratio]{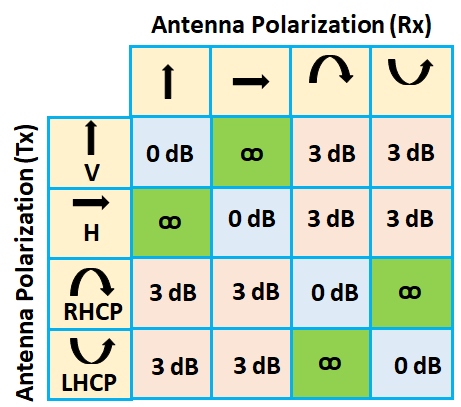}
     \vspace{-0.2cm}
     \caption{A conceptual illustration of the various possibilities of polarization misalignment and subsequent mismatch loss. Polarization is defined by the orientation of the maximum electric field vector. The effect of polarization misalignment between an incoming electromagnetic signal and a receive antenna may render links inoperable in the most extreme cases.}
     \label{loss}
 \end{figure}

\begin{figure}[!t]
 \centering
     \includegraphics[width=\columnwidth]
     {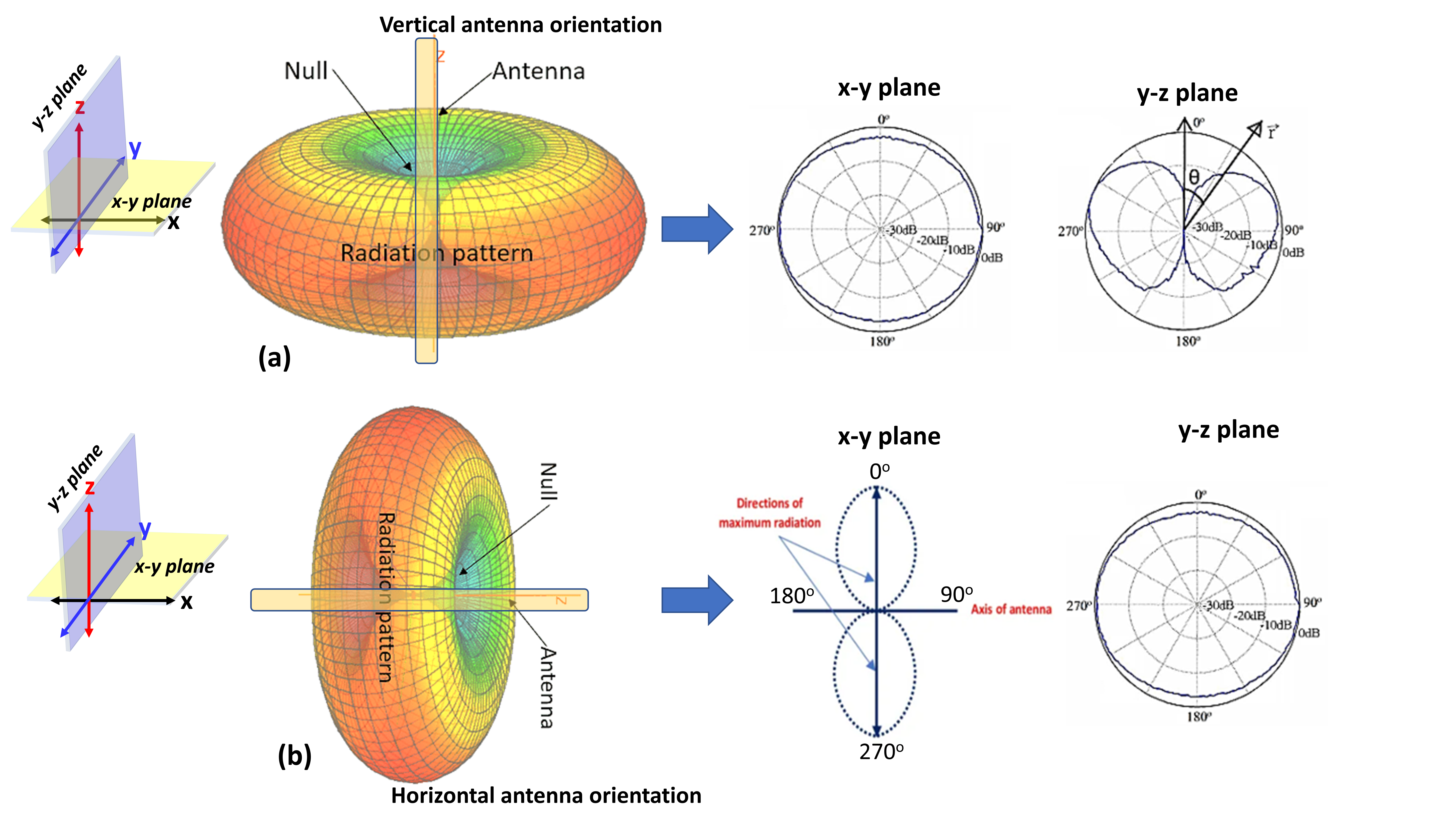}
     \vspace{-0.2cm}
     \caption{Tx and Rx antenna radiation patterns and orientations.}
    \label{pattern}
 \end{figure}

The antenna polarization is also an important parameter to be considered in communication systems. For instance, if the TX antenna on a UAV is vertically polarized (V) while the user end is horizontally (H) polarization (or vice versa), the communication link will not be established even in the LOS scenario. There will be a significant degradation in received signal strength due to polarization mismatch if the antenna orientations are not aligned properly, even when the drone is near the ground receiver \cite{U4}. Therefore, the correct polarization alignment of antennas is crucial. A demonstration of the loss in the received power due to polarization mismatch is presented in Fig. \ref{loss}. For V-H antenna alignment, the signal is completely lost. Maximum signal strength is achieved for V-V or H-H alignment (0 dB shows maximum received power or equivalently the least power loss). The loss of $3$dB (or the equivalent in the other direction) is experienced when attempting to receive a linearly polarised signal with a circularly polarised antenna, although they can typically be handled. The use of an orthogonal antenna polarization poses the most significant loss in power because the attenuation exceeds all theoretical bounds. Since most antennas only have a little amount of polarization decoupling, in reality, the loss will never be infinite practically. 
We provide an analytical model for the case of a dipole (or a monopole) antenna for UAV to ground terminal connection. The radiation pattern for both antennas is similar; therefore, the same mathematical model can serve in both cases. Hence the dipole antenna will be considered onward for analysis. A dipole antenna has a doughnut-shaped radiation pattern in 3D or an 8-shaped pattern in 2D view in E-plane. The antenna radiation pattern is critical to understand because the LOS scenario is directly concerned with antenna connectivity in a particular direction. The 3D radiation patterns are also analyzed in 2D, which is shown in Fig. \ref{pattern}. Based on the position of the dipole, we can have either a horizontal cut (x-y plane) or an elevation cut (y-z or x-z planes or E-plane). In \cite{U4}, a sine function was mapped with the elevation gain of the antenna because, in that scenario, the UAV was hovering while the user (RX) remained static. In that case, the bore-sight antenna direction, $\theta=0\degree$, provides null because $\sin(\theta)$ or $\cos(90\degree-\theta)$ is zero, while the maximum is obtained at 90\degree and 270\degree directions. In contrast, this work considers a moving user, i.e. car, while the UAV is assumed to be static. This scenario requires maximum gain at LOS, for which the radiation pattern of the antenna should have maximum directivity at 0 \degree and 180\degree, respectively. This can be achieved if the dipole antenna is placed horizontally, as shown in Fig. \ref{pattern} (b). In this alignment, the horizontal (x-y) plane will serve as a directional plane with an 8-shaped pattern in 2D, while an omnidirectional pattern (a circle in 2D) will be obtained in the vertical (y-z) plane.

\begin{figure}[!t]
 \centering
     \includegraphics[width=\columnwidth]{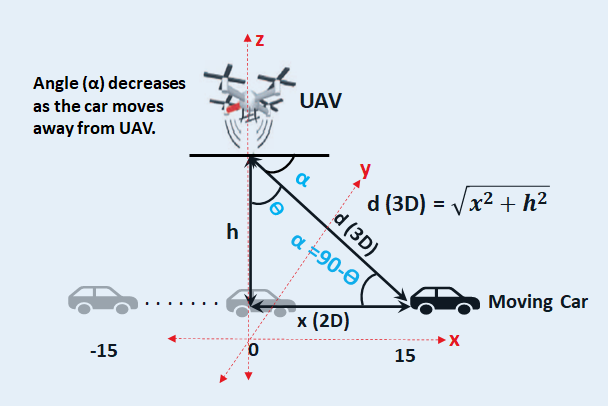}
     \vspace{-0.2cm}
     \caption{A conceptual modeling of UAV and moving car in 2D.}
     \label{angle}
 \end{figure}

According to the well-known Friis' transmission equation (in linear scale):
\begin{equation}\label{eq1}
P_{Rx} = P_{Tx} \times G_{Tx}  (\alpha) \times G_{Rx}(\alpha) \times  {(\frac{4\pi d} {\lambda})}^\gamma 
\end{equation}
where $\gamma$ is the path loss exponent, $\alpha$ is the angle between the Tx and Rx as shown in Fig. \ref{angle}. The angle of elevation $\alpha$ between Tx (UAV) and Rx (car) increases when the horizontal distance between the car and the UAV decreases. The moment when UAV and car are perpendicular, this angle $\alpha$ becomes maximum and nulls are aligned with each other (VV alignment), thus connectivity is lost for VV orientation. Hence, we need to employ HH to achieve maximum gain at LOS direction when UAV and car are perpendicular i.e., (when $\alpha$ = 90\degree, and $\sin(90\degree) = 1$ or $\cos(0)=1$, because of $\sin(\alpha)=\cos(90\degree-\alpha)$.

\textbf{Received Signal Modelling for LOS Scenario}\\
For a transmitted signal T(n), the received signal at the receiver antenna can be written as the convolution of the transmitted signal with the channel impulse response as \cite{U5}:
\begin{equation}
    R(n) = T(n) * H(n)
\end{equation}
where $H(n)$ is the impulse response of the channel. A generic representation of the received signal as a combination of all multi-path signals can be represented as:
\begin{equation}
\begin{split}
R_{i} (n) = & \frac{\lambda \Gamma_{i} (\theta, \phi)}{4 \pi d_{i}} * \sqrt{G_{T}(\phi_{i}^{(TX)}, \theta_{i}^{(TX)}) G_{R}(\phi_{i}^{(RX)}, \theta_{i}^{(RX)})} \\ 
& s(n - \tau_{i}) \exp(\frac{-j 2 \phi d_{m}}{\lambda} \lvert \psi_{i}^{TX} . \psi_{i}^{RX} \rvert),
\end{split}
\end{equation}

where $i=0,1,2…$ is the multi-path component, $\Gamma_{i}$ is the reflection coefficient of $i_{th}$ multi-path component, $\phi_{i}$ is the polarization mismatch loss factor, $\theta$ and $\phi$ are the elevation and azimuth angles between Tx and Rx, $d$ is the distance between Tx and Rx, and $\tau_{i}$ is the delay of $i_{th}$ multipath component.

It is instructive to mention here that some assumptions are made to make our scenario simple: the path loss exponent $\gamma$  given in (\ref{eq1}) is assumed to be 2, for the LOS scenario. The polarization mismatch is assumed to be $1$ (0 dB) and multipath reflections or ground reflections are not considered for UAV channel modeling. As a result of these assumptions, where for LOS case, $i=0, \Gamma=1, \tau_{o}=0, and 
 \phi_{o}=1,$ while $G_{Tx}$ and $R_{Tx}$ can be mapped to sinusoidal function, we are left with a simplistic mathematical formulation for LOS component of the received signal as:

\begin{equation}
    R_{o} (n) = \frac{\lambda}{4 \pi d_{o}} * \sin(\theta, \phi) * T(n).
\end{equation}

Since the power of a signal can be obtained by its modulus square, thus $P_{tx} = | T(n)|^2$ and the received power for the LOS component can be modeled as:

\begin{equation}
    P_{RX} =  \frac{P_{TX} \sin^2(\theta)\lambda^2}{(4\pi d_{o})^2},
\end{equation}
where, $\theta$ is the elevation angle between the UAV and the car which can be found as $(\theta = \tan^{-1}(x/d))$ which decreases as the car moves away from the UAV, or if the height of the UAV increased (note that angle $\theta$ and $\alpha$ respond oppositely, and we are interested in $\alpha$). The moment the car is in front of the UAV, $\theta = 0\degree$, and maximum received power is obtained. However, as the car becomes in line with the UAV, the received power should become maximum, which can be achieved if the angle for the sinusoidal function is 90\degree, i.e.,  $\alpha = 90\degree$. This can be achieved if the actual angle $\alpha$ is used in the equation, and the correct input trigonometric angle in (5) will then become $\alpha=90\degree-\theta\degree$. As a result, $\sin(90\degree-0\degree)=\sin(90\degree)=\cos(0\degree) $$= 1$, showing maximum RSSI.

\section{UAV-assisted Handover Mechanism} 
To predict future beam blockage, CV and DL are utilized whereas UAV is considered as a BS for proactive HO. In a realistic wireless communication scenario, future blockage prediction is challenging as it depends on the user's speed and surrounding environment. Our previous study used bimodal data (vision and wireless) for beam blockage prediction, which assisted in triggering the optimal HO \cite{viwi1}. The CV-aided blockage prediction is divided into two sub-tasks, i.e., (i) ODL to determine the location and type of the blocking object from the RGB image to calculate the user's speed. (ii) Predict $(T_{blk})$ using the information extracted for RGB images. 

In the proposed setup, if the user and blocking object are in the field of view (FoV), it triggers the blockage (BLK) event. The key idea is to obtain the location of the user and blocking object once the BLK event is detected. The location information and speed of the user are used to determine the $T_{blk}$, which allow us to perform proactive HO before service disconnection. The details of the proactive HO mechanism are discussed in our previous study \cite{9926150}. However, the brief steps involved in HO are highlighted as: 
\begin{itemize}
    \item The ODL algorithm detects the BLK event and provides augmented information on user location and speed of the user.
    \item The information extracted by ODL is used to predict the $T_{blk}$.
    \item In the final stage, the CU performs the HO if the time to execute HO ($T_{exec}$) is greater than the $T_{blk}$.
\end{itemize}

The $T_{exec}$ is the minimum time required for the proposed scheme to perform successful HO, which is divided into four-sub times and mathematically given as \cite{9926150}:
\begin{equation} \label{eq22}
    T_{exec} = T_{RGB} + T_{ODL} + T_{inf} + T_{HO},
\end{equation}
where $T_{RGB}$ is the time required to transmit RGB images to CU, $T_{ODL}$ for object detection and localization, $T_{inf}$ is the inference time for the regression model, and $T_{HO}$ is the time to perform HO. Furthermore, a new time parameter, time waiting ($T_{w}$) is defined, with the maximum value as the difference between $T_{blk}$ and $T_{exec}$ represented mathematically as:

\begin{equation}
    T_{w} ^{max} = T_{blk} - T_{exec}
\end{equation}

It is worth mentioning that all the parameters given in (\ref{eq22}) are fixed. However, $T_{w}$ is dependent on the $T_{blk}$, which is the function of user location and speed. Before discussing the details of each component, the following are the safe assumptions made in this study:
\begin{itemize}
    \item The SBSs have both vision and wireless sensor data and continuously send it to the CU for real-time inference.
    \item All the data processing is done on the CU, and the ODL model correctly identifies the coordinates of the potential blocking object and $T_{blk}$ is also estimated correctly.
    \item The UAV is optimally placed at a certain height, covering the entire blockage area.
    \item The $T_{blk}$ is greater than the $T_{exec}$, resulting enough $T_{w}$ to perform HO successfully.   
\end{itemize}

\subsection{ODL and Time to Block Prediction}
In this HO mechanism, ODL plays a key role as it provides the accurate location of the user and potential blocking object. The obtained location coordinates are used to determine the user's speed. The details on the ODL to determine the blockage event are beyond the scope of this paper as they are already discussed in our previous study \cite{9926150}. However, very briefly, the ODL is divided into the following two sub-tasks: (i) A pre-trained you only look once (YOLO) version 3 (YOLOv3) model is used to obtain the pixel coordinates of the user and potential blocking object. (ii) The 2D-pixel coordinates obtained by YOLOv3 are transformed into displacement coordinates to determine the user's speed. Since the optimal HO depends on the $T\_{exec}$, we need to find all times given in the (\ref{eq22}) to perform a successful HO. 

Once the location coordinates and user speed are obtained using ODL, $T_{blk}$ is predicted using a pre-trained neural network model. For initial model training, the dataset is extracted for the information-rich RGB images, which includes the location coordinates and speed of the user. For simplicity, the location of the blocking object is kept fixed with varying user location and user speed. The model training is done offline, however, in our proposed scheme, real-time inference is done to obtain the $T_{blk}$. Based on our analysis in \cite{9926150}, the $T_{ODL}$ is approximately $102$ ms, whereas the $T_{inf}$ is around $1$ ms.

\subsection{Optimal Trigger Region and Final Handover}
The parameters like user speed, location coordinates, and  $T_{blk}$ are used to determine the optimal distance to perform HO. The optimal trigger distance "D" is determined by performing an in-depth analysis using the threshold distance-based setup given by the following equation:
\begin{equation}
    D \leq S_{u}(T_{blk} - T_{exec}),
\end{equation}
where $S_{u}$ is the user speed known, $T_{exec}$ is the sum of four sub-times given in (\ref{eq22}). The analysis is done based on different user speeds to determine the $T_{w}$ \cite{9926150}. Furthermore, the impact of early HO on the QoS is also studied. The final stage in our proposed scheme is the HO, in which $T_{HO}$ parameter determines whether the successful HO is possible or not. For successful HO, $T_{blk}$ should be greater than $T_{exec}$, ensuring enough $T_{w}$. In our case, the $T_{exec}$ is approximately $153$ ms, and if the CU detects the BLK event, the $T_{blk}$ should be greater than the $153$ ms for a successful HO. In the worst-case scenario, the user will face connection failure if the $T_{blk}$ is less than $T_{exec}$. 

\begin{algorithm}[H]
\SetAlgoLined
\KwResult{Handover decision and execution}
Initialisation\;
BLK $\leftarrow$ False\;
Initialise $S_u$, $L_u$\;
\While{True}{
  BLK, $L_u$, $S_u$ $\leftarrow$ ODL\_Module()\;
  \eIf{BLK}{
   $T_{blk} \leftarrow T_{blk}(L_u, S_u)$\;
   $T_{exec} \leftarrow T_{exec}()$\;
   \If{$T_{blk} > T_{exec}$}{
    $RSSI_{BS} \leftarrow RSSI_{BS}()$\;
    $RSSI_{UAV} \leftarrow RSSI_{UAV}()$\;
    \eIf{$RSSI_{UAV} > RSSI_{curr}$}{
     Switch connection to UAV\;
     }{
     Switch to the BS\;
    }
   }
   }{
   Continue Monitoring (No Handover required)\;
  }
 }
\caption{Proactive UAV-Assisted Handover Algorithm}
\end{algorithm}
The following is the process that the algorithm 1 follows to determine when to switch the connection between a user's device and a Base Station (BS) or an Unmanned Aerial Vehicle (UAV).
\begin{enumerate}

 \item  Initialization: The algorithm starts by setting the blockage status (BLK) to False, and initializing the user's speed ($S_{u}$) and location ($L_u$).

 \item  Continuous Monitoring: The algorithm runs continuously in a loop, keeping an eye on the network status.

 \item   ODL\_Module: A function that updates the BLK status, user location, and speed.

 \item  Handover Decision: 
If blockage is detected (BLK == True), the algorithm calculates $T_{blk}$ and $T_{exec}$. If $T_{blk}$ is greater than $T_{exec}$, the algorithm proceeds to check the Received Signal Strength Indicator (RSSI) from both the Base Station (BS) and the UAV. Based on the RSSI values, the algorithm decides whether to switch the connection to the UAV or another BS.

 \item  No Handover: If there is no blockage detected, the algorithm continues to monitor without initiating a handover.
\end{enumerate}

\section{Simulation and Results} 

For simulation and performance analysis of UAV-assisted HO, we used a publically available vision-aided wireless communication (ViWi) dataset \cite{viwi1}. ViWi is a parametric and scalable dataset generation platform that combines visual and wireless data information. The ViWi platform generates the high-fidelity synthetic dataset using Wireless InSite software (ray tracing for visual data) and a 3D game (a blender for visual data). This dataset considers multiple scenarios based on the placement of visual sensors and the user's view. The camera placement considers distributed (placed on multiple base stations) or co-located (placed on a single base station) scenarios. In the case of the user's view, two scenarios (direct and blocked) are considered. Since this work focuses on proactive blockage prediction; therefore, we merged the two scenarios, i.e., co-located camera direct view and co-located camera with a blocked view. The idea of merging the two scenarios is to parameterize the dataset for the blockage prediction problem.  

\subsection{Simulation Setup}
 The simulation setup considers a simple scenario: a single user, a blocking object, and a UAV placed above the blocking area, as shown in Fig. \ref{sm}. The user is moving from right to left and being served by $SBS_{1}$. However, a potential blocking object may interrupt the Line of Sight (LoS) communication, leading to service disconnection. Therefore, proactive handover (HO) is necessary to eliminate service disconnection. In our previous study, a successful HO is performed by proactively shifting the user from $SBS_{1}$ to $SBS_{2}$ for seamless connectivity. However, the HO resulted in the degradation of RSSI, which is undesirable. Hence, to overcome this challenge, we propose a UAV-assisted HO mechanism to maintain seamless connectivity with minimum performance degradation.
 
 The initial challenge was the availability of UAV channel characteristics with respect to moving users. To generate the UAV data samples, we merged the two scenarios of the ViWi dataset. Using the co-located camera with the direct view, we plotted the RSSI based on the user location. Using the UAV model analysis discussed in Section \ref{uav_channel}, we provide different values of RSSI for various heights, such as 15m, 20m, 25m, and 30m, for Free Space Path Loss
(FSPL) model and the Two-Ray Ground Reflection model  are shown in Fig. 6 \& 7. 
 
 \subsection{Impact of Channel Model on RSSI}

 Since UAVs operate at varying altitudes, it is imperative to understand how communication performance is impacted by signal propagation. The use to various channel models, which predict radio waves behavior in different scenarios, is the fundamental basis for this understanding. This paper presents a comparative analysis of the Received Signal Strength Indicator (RSSI) results obtained from two commonly used channel modeling techniques: the Two-Ray Ground Reflection model and the Free Space Path Loss (FSPL) model \cite{chiu2021channel}, at different UAV heights. 
 
 The analysis primarily focuses on how these models anticipate the behavior of communication signals, which vary with UAV height and horizontal distance from the ground user in UAV-to-ground user scenarios. The Free Space Path Loss (FSPL) model and the Two-Ray Ground Reflection model are two popular models that offer fundamental understandings of the physics of radiofrequency communication in UAV applications. The FSPL model is especially simple to use and important for determining the fundamental attenuation of radio waves with distance and frequency as it assumes a clear line-of-sight (LOS) across the receiver and the transmitter. When assessing UAV communications in high-altitude situations where the LOS component is dominant, this model becomes an essential tool. 
 \begin{figure}[!t]
 \centering
     \includegraphics[width=\columnwidth]{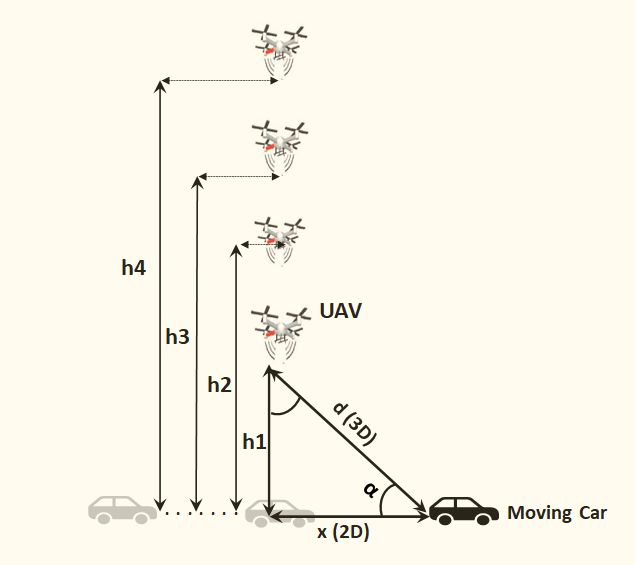}
 
     \caption{Illustration of varying heights of UAV with respect to moving car.}
 \end{figure}
\begin{figure*}[h!]
  \centering
  \begin{minipage}{0.48\textwidth}
    \centering
    \includegraphics[width=\linewidth]{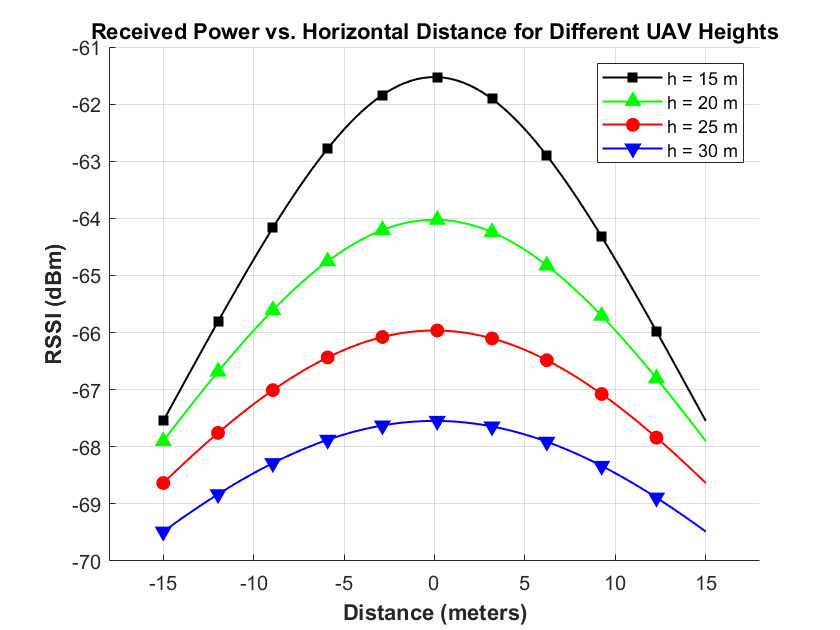}
    \caption{Effect of varying UAV heights on RSSI for Free Space Path Loss
(FSPL) model. The curve for h=20 m is used for the results for a co-located camera with a direct view for proactive HO. }
    \label{figrssi}
  \end{minipage}\hfill
  \begin{minipage}{0.48\textwidth}
    \centering
    \includegraphics[width=\linewidth]{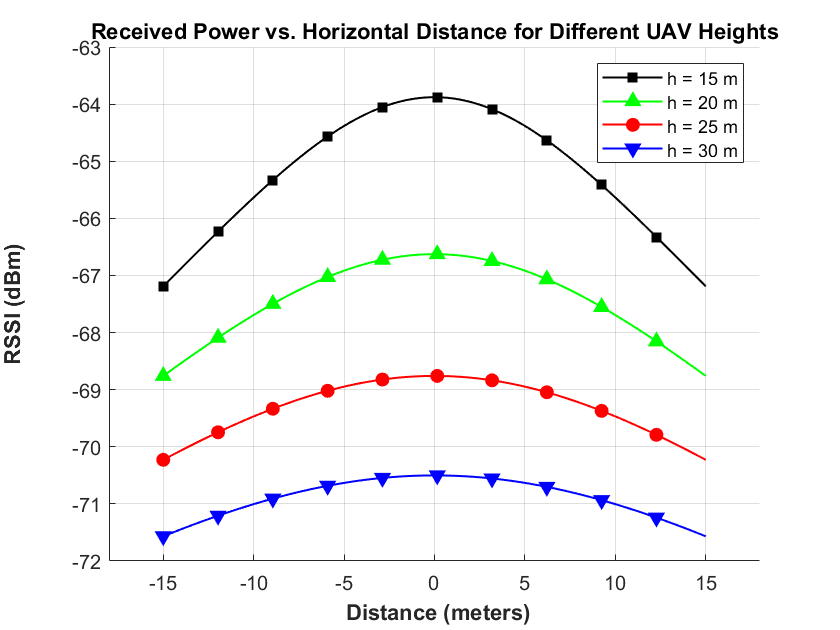}
    \caption{Effect of varying UAV heights on RSSI for Two-Ray Ground Reflection model  Model. The curve for h=20 m is used for the results for a co-located camera with the direct view for proactive HO.}
  \label{fig:sidebyside}
  \end{minipage}
  
\end{figure*}

The path loss \( L \) in decibels is given by:

\begin{equation}
L = 20 \log_{10}(d) + 20 \log_{10}(f) + 20 \log_{10}\left(\frac{4\pi}{c}\right)
\end{equation}

where:
\begin{itemize}
  \item \( d \) = Distance between the transmitter and receiver
  \item \( f \) = Frequency of the signal
  \item \( c \) = Speed of light
\end{itemize}

In non-logarithmic form, the received power \( P_r \) is:

\begin{equation}
P_r = \frac{P_t G_t G_r \lambda^2}{(4\pi d)^2}
\end{equation}

where:
\begin{itemize}
  \item \( \lambda \) = Wavelength of the signal ( \( \lambda = \frac{c}{f} \) )
\end{itemize}
 However, for low-altitude UAV operations, a more thorough analysis is provided by the Two-Ray Ground Reflection model, which takes into account the impacts of both a direct path and a ground-reflected path. Understanding how ground reflections can improve or degrade the signal based on the phase difference between the direct and reflected channels is crucial for understanding the links between the transmitted signal and the surrounding environment.

The received power \( P_r \) is given by:

\begin{equation}
P_r = \frac{P_t G_t G_r h_t^2 h_r^2}{d^4}
\end{equation}

where:
\begin{itemize}
  \item \( P_t \) = Transmitter power
  \item \( G_t \) and \( G_r \) = Gains of the transmitter and receiver antennas, respectively
  \item \( h_t \) and \( h_r \) = Heights of the transmitter and receiver antennas, respectively
  \item \( d \) = Distance between the transmitter and receiver
\end{itemize}

The phase difference between the direct and reflected paths can be expressed as:

\begin{equation}
\Delta \phi = \frac{2\pi d}{\lambda}
\end{equation}

where \( \lambda \) is the wavelength of the signal.

 The RSSI value decreases for FSPL model from -61.5 dBm to -67.5 dBm  and for the two-Ray Ground Reflection model the RSSI decreases from -64 dBm to -71 dBm as the height increases from 15m to 30m. This decrease occurs because the antenna gain and coverage beamwidth are affected, along with an increase in path loss, as the height of the UAV increases. Similarly, RSSI drops as the horizontal distance between the UAV and the car increases because of the inverse square variation of path loss with that of the distance. After the detailed analysis, the height of 20m replicated the results for the ViWi scenario with the smooth bell-shaped curve of RSSI. Therefore, the RSSI of 20m height is used as a dataset for UAV to perform HO once the $BLK$ event is detected.

\subsection{Results and Discussion}
The proposed UAV-assisted HO mechanism maintains seamless connectivity with minimum performance degradation. The performance of the user during Handover (HO) is assessed using the normalized Received Signal Strength Indicator (RSSI) metric. While the user is initially served by $SBS_{1}$, service disconnection occurs in front of a blocking object. Upon detecting a Blockage ($BLK$) event, the HO algorithm determines the time needed for handover, denoted as $T_{blk}$. If the execution time, $T_{exec}$, is less than $T_{blk}$, the HO request is initiated. The final HO is carefully executed, considering an optimal trigger region. In this scenario, a UAV is assumed to be positioned at a specific height, providing coverage to the blockage area.
\begin{figure}[h!]
	\centerline{\includegraphics[width=\columnwidth]{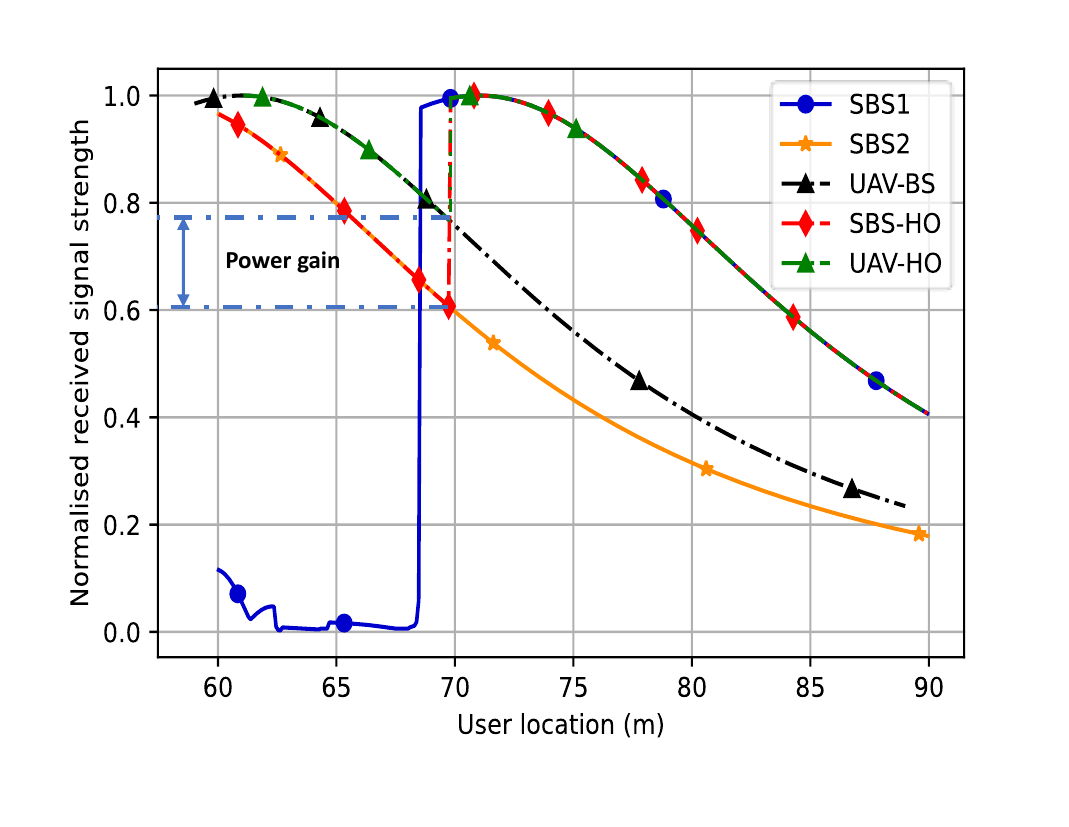}}
	\caption{Results of UAV assisted HO. The user is served by $SBS_{1}$ and experiences service disconnection without HO.}
	\label{results}
\end{figure}

The results of the final HO are depicted in Fig. \ref{results}. Notably, with the optimal HO of the user from $SBS_{1}$ to $SBS_{2}$, a significant degradation in RSSI is observed, which is undesirable. This drop in RSSI is attributed to path loss, as $SBS_{2}$ is relatively far from the user.

In our proposed solution, a UAV is employed as a BS for HO. In the context of UAV-BS, a drop in the Received Signal Strength Indicator (RSSI) of the user occurs due to path loss. However, despite this, the overall performance of UAV-HO surpasses that of HO without UAV involvement. To clarify, at the optimal trigger distance, UAV-HO yields a $20\%$ increase in RSSI compared to HO scenarios where a UAV is not utilized, as illustrated in Fig. \ref{results}. This highlights the effectiveness of incorporating a UAV as a BS in improving the user experience during Handover.

\subsection{Quality of Experience (QoE)}
\begin{figure}[h!]
	\centerline{\includegraphics[width=\columnwidth]{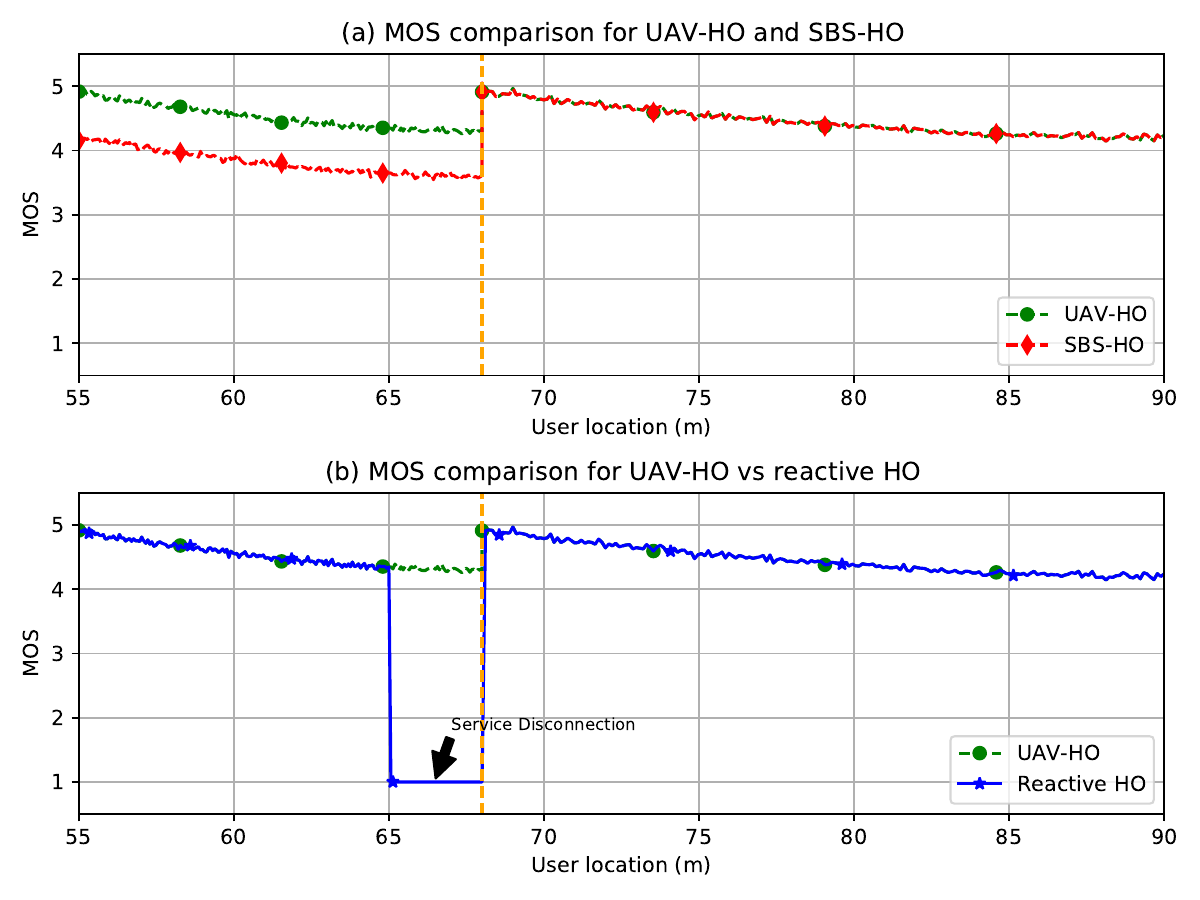}}
	\caption{Comparison of UAV-HO, SBS-HO and reactive, measured by MOS. The users have a lower MOS when shifted to $SBS_{1}$ and experiences service disconnection for reactive H0.}
	\label{mos}
\end{figure}

In this section, we discuss how the PHO algorithm contributes to improving the reliability of high-frequency wireless networks as perceived by a real-time application that is, in turn, sensitive to service interruptions and network latency. A working example is the mobile user who is video-calling, and the metric is the Mean Opinion Score (MOS). The MOS evaluates the user's quality of experience (QoE) and is rated by the human perception of the overall quality of the service, with scores between 1 and 5 (1: bad, 2: poor, 3: fair, 4: good, and 5: excellent) \cite{9926150}. The RSS values are mapped to the corresponding MOS by using the table given in \cite{mkwawa2012mapping}. The results in Fig. \ref{mos} show the MOS  comparison for both proactive and reactive HO. For instance, in the case of PHO, Fig. \ref{mos}(a) shows the MOS for UAV-HO and SBS-HO. UAV-HO maintains a higher level of RSS value, resulting in high MOS ranging between 4 and 5. However, in the case of SBS-HO, the MOS is reduced below 4 when the HO happens, which causes a lowering in the user experience. Furthermore, in the case of reactive HO, there will be a service disconnection, as given in Fig. \ref{mos} (b), until it establishes the re-connection.    
\section{Conclusion and Future Work} 
 
This paper introduces a Computer Vision (CV)--based proactive Handover (HO) strategy designed to mitigate blockages, with a UAV serving as a Base Station (BS). The central concept of this approach involves enhancing the awareness of the wireless network about its environment through the utilization of multi-modal data. A primary focus is placed on proactive blockage prediction, a crucial application that facilitates HO to guarantee seamless connectivity. In terms of antenna modelling on the UAV end, a polarization-matched scenario is elucidated to optimize the received signal. The results indicate that with UAV-assisted HO, users not only maintain seamless connectivity but also witness an overall performance improvement of $20 \%$.

In our future work, the idea is to enhance the UAV channel modelling with respect to reflections and multipath channels. Furthermore, obstructions due to foliage and buildings need to be modelled to provide a robust UAV channel model such that a constructive interference scheme of these multiple paths can be devised to enhance the gain and coverage quality on the user side.

\bibliographystyle{elsarticle-num} 
\bibliography{bib}






\end{document}